\documentclass[aps,pra,10pt,twocolumn,floatfix,nofootinbib]{revtex4-1}

\usepackage{bbm}
\usepackage{amsmath}
\usepackage{amssymb}
\usepackage{graphicx}
\usepackage{amsfonts}
\usepackage{amsthm}

\newtheorem{thm}{Theorem}[section]
\newtheorem{cor}[thm]{Corollary}
\newtheorem{lem}[thm]{Lemma}
\newtheorem{prop}[thm]{Proposition}

\theoremstyle{definition}
\newtheorem{defn}[thm]{Definition}
\newtheorem*{assump1}{Infinite reducibility (or classical) assumption}
\newtheorem*{assump2}{Determinism and Reversibility assumption}
\newtheorem*{assump3}{Kinematic equivalence assumption}

\begin{document}

\title{From physical principles to relativistic classical Hamiltonian and Lagrangian particle mechanics}
\author{Gabriele Carcassi}
\affiliation{University of Michigan, Ann Arbor, MI 48109}
\email{carcassi@umich.edu}
\date{March 15, 2015}

\begin{abstract}
We show that classical particle mechanics (Hamiltonian and Lagrangian consistent with relativistic electromagnetism) can be derived from three fundamental assumptions: infinite reducibility, deterministic and reversible evolution, and kinematic equivalence. The core idea is that deterministic and reversible systems preserve the cardinality of a set of states, which puts considerable constraints on the equations of motion. This perspective links different concepts from different branches of math and physics (e.g. cardinality of a set, cotangent bundle for phase space, Hamiltonian flow, locally Minkowskian space-time manifold), providing new insights. The derivation strives to use definitions and mathematical concepts compatible with future extensions to field theories and quantum mechanics.\end{abstract}
\maketitle

\section{Introduction}

Classical particle mechanics is usually founded on Newton's laws. These, though, are insufficient to derive the full Lagrangian and Hamiltonian formalism, and usually other ad-hoc assumptions (e.g. conservative forces) are introduced. Special relativity is based on two principles (invariance of the speed of light and the principle of relativity), which lead to the Minkowskian nature of space-time but not to the equations of motion, which are a consistent reformulation of the non-relativistic ones. What we'll do in this work is re-organize all the known elements and equations in a more consistent and comprehensive way, leading to better insight on why the fundamental concepts and laws are what they are. The derivation will also include classical analogues of strictly quantum notions to facilitate a future extension to quantum mechanics.

We'll use concepts from different disciplines, such as set theory, differential geometry, relativity, Hamiltonian and Lagrangian mechanics, and we'll find interesting connections among them. We'll keep names and notation as consistent as possible to current use across the different disciplines. This may sometimes lead to some non sequitur as it will not be immediately clear why the new definitions are equivalent to the standard ones. These are typically resolved by subsequent derivation of the expected properties.

No mathematical breakthrough should be expected: the goal, after all, is to derive the \emph{known} framework from a set of \emph{simple} definitions in the most \emph{obvious} way possible. No proof is longer than a couple of paragraphs, so the word \emph{theorem} is avoided in favor of \emph{proposition} and \emph{corollary}. The novel, and surprising, result is how so much can be derived from so little.

\section{Outline}

We give here a brief general conceptual overview, hoping it will help guide the reader through the mathematical details.

We'll first mathematically define states and the labels we use to identify them (e.g. position, momentum, temperature, pressure). We introduce the infinite reducibility assumption (or classical assumption): each state is divisible into the states of its parts. It is then sufficient to describe the evolution of infinitesimal parts (i.e. particles) within their state space (i.e. phase space).

We introduce the deterministic and reversible evolution assumption: to each state corresponds one and only one future (or past) state. The cardinality of a set of states is therefore conserved under evolution. This allows us to define a metric $\omega$ on phase space, and the conservation of that metric leads to the Hamiltonian framework. That is: we can derive the Hamiltonian framework on its own merit and show that it is equivalent to deterministic and reversible motion.

We introduce the assumption of kinematic equivalence: studying trajectories is equivalent to studying states. We show that there must be a link between the space-time metric $g$ and the phase-space metric $\omega$, as both quantities must be conserved under passive coordinate transformations. That link will constrain the space-time metric to be locally Minkowskian. A transformation between state variables $(q, p)$ and kinematic variables $(x, \dot{x})$ must exist.  To be invertible, the relationship between velocity and conjugate momentum has to be monotonic, leading to a concave Hamiltonian which allows a Legendre transformation leading to the Lagrangian. We can constrain the Hamiltonian further, and show that the most general equation of motion under the three assumptions is the one of a geodesic modified by the force given by a vector potential (such as the one of a relativistic charged particle) and a scalar potential (such as the one of Newtonian gravitation in the non-relativistic case).

\section{States, Labels and Maps}

This and the next section are dedicated to properly defining states. In particular, we will need to introduce more precise terminology to be able to make two crucial distinctions. The first is between the state of the whole system and one of its parts.\footnote{Whole system vs. infinitesimal component is also a source of confusion when comparing classical and quantum states. Quantum states are always whole and are always distributions as there is no state attributable to parts of a quantum system.}
\begin{center}
    \begin{tabular}{ | p{2.5cm} | p{5.5cm} | }
    \hline
    Configuration state & The state of the whole system being described. \\ \hline
    \raggedright{Particle state (or simply state)} & The state of an infinitesimal part (i.e. particle) of the system. \\ \hline
    \end{tabular}
\end{center}
When talking about particles of a system we must always remember that they are the result of a limit. Therefore we will start with discrete definitions, and see how some properties will extend to the continuous case.

The second distinction comes up when dealing with physical quantities. We need to be able to distinguish among a particular value, a set of possible values, and the space of all possible values.
\begin{center}
    \begin{tabular}{ | p{2.5cm} | p{5.5cm} | }
    \hline
    State variable & A quantity that must be specified to identify a state (e.g. position). \\ \hline
    Label & A particular value for a state variable (e.g. position = $5m$). \\ \hline
    Label set & A set of possible values for a state variable (e.g. position = $[4.5m, 5.5m]$). \\
    \hline
    \end{tabular}
\end{center}

We will first define discrete labels using set theory and in the next section generalize to the continuous case using differential geometry. These new terms are fundamental as most of the later derivation will be based on \emph{counting the labels} (i.e. counting the possible values of a physical quantity) and making sure that such number is conserved (i.e. the number of cases is the same).

\begin{defn}\label{statedef}
Fix a physical system to study. We define the set $\mathbbm{C}$ of all physically distinguishable configurations for that system. Each element $\mathbbm{c}$ we call \emph{configuration state}.
\end{defn}

\begin{assump1}\label{classical}
The system is infinitely reducible: it can be thought of as composed of two or more similar but smaller systems, each in its own configuration state, which in turn can be thought of as composed of two or more, ad infinitum. We call \emph{particle} such an infinitesimal part.
\end{assump1}

\begin{defn}\label{classicalPhaseSpace}
Let $\mathbbm{S}$ be the set of all possible configuration states for a particle. We call this set \emph{phase space}. We call each $\mathbbm{s} \in \mathbbm{S}$ a \emph{particle state}, or simply \emph{state}.
\end{defn}

\begin{cor}\label{classicalDistribution}
Each classical configuration state $\mathbbm{c} \in \mathbbm{C}$ is a distribution over particle states: $\mathbbm{c}=\sum\limits_{\mathbbm{s} \in \mathbbm{S}} D(\mathbbm{s}) \mathbbm{s}$, where $D:\mathbbm{S}\rightarrow\mathbb{R}$ measures how much of the system can be found in each $\mathbbm{s}$. The distribution can be visualized as a histogram over the states in phase space.
\end{cor}

Under the classical assumption, we can then limit ourselves to study the particles of the system, their states and their properties without loss of generality. To help identify states, we introduce the following concepts.

\begin{defn}\label{label}
We call a \emph{label} a set of states $i\subset\mathbbm{S}$; a \emph{set of labels} a collection of disjoint labels $I | \forall i_1,i_2\in I, i_1\bigcap i_2 = \emptyset$; a \emph{state variable} a set of labels $\mathbbm{I}$ that covers all of phase space: $\bigcup\limits_{i \in \mathbbm{I}}i=\mathbbm{S}$. Therefore a state belongs to one and only one label of a state variable.
\end{defn}

\begin{defn}\label{discreteCardinality}
Let $I$ be a finite, countable set of labels. We define the \emph{cardinality} $n(I) \rightarrow \mathbbm{N}$ as the number of labels in the set.
\end{defn}

\begin{defn}\label{labelsCombine}
Let $I_1$ and $I_2$ be two sets of labels. We can define the \emph{combined set}, $\langle I_1, I_2 \rangle$, whose labels consist of all the non-empty intersections of one label of $I_1$ and one of $I_2$. If all intersections are non-empty, $I_1$ and $I_2$ are said to be \emph{independent}, and we have $n(\langle I_1, I_2 \rangle)=n(I_1)n(I_2)$.
\end{defn}

We now want to study how states and labels evolve in time, under the following assumption.

\begin{assump2}
The system undergoes deterministic (future state identified by the present state) and reversible (past state identified by the present state) evolution.
\end{assump2}

\begin{prop}\label{detrevMap}
Let $\mathbbm{S}$ be the phase space of a system that undergoes deterministic and reversible evolution. There exists a bijective map $f:\mathbbm{S} \leftrightarrow \mathbbm{S}$ between past and future states.
\end{prop}

\begin{cor}\label{detrevDist}
The evolution of a classical configuration state $\mathbbm{c}=\sum D(\mathbbm{s}) \mathbbm{s}$ under a bijective map is given by $\mathbbm{c'}=\sum D'(\mathbbm{s}) \mathbbm{s}=\sum D(f^{-1}(\mathbbm{s})) \mathbbm{s}$. The evolution of the fraction of the system in a label $D(i)=\sum\limits_{\mathbbm{S} \in i} D(\mathbbm{s})$ is given by $D'(i)=D(f^{-1}(i))$.
\end{cor}

Mathematically, assuming determinism and reversibility means studying bijective maps. The evolution of a distribution simply moves the elements around: the bars of the histogram move place, but keep the same height.

\begin{cor}\label{labelsCount}
Given a label $i$, the image $f(i)$ is also a label containing the same number of states. Given a set of labels $I$, the image $f(I)$ is also a set of labels containing the same number of labels $n(I) = n(f(I))$. Given a state variable $\mathbbm{I}$, the image $f(\mathbbm{I})$ is also a state variable. Given two independent sets of labels $I_1$ and $I_2$, the images $f(I_1)$ and $f(I_2)$ are also independent. Therefore $n(f(\langle I_1, I_2 \rangle))=n(f(I_1))n(f(I_2))=n(I_1)n(I_2)=n(\langle I_1, I_2 \rangle)$
\end{cor}

Bijective maps preserve the number of labels as they provide one-to-one association between future and past. And they do so for each independent set of labels. These simple results using discrete labels, properly generalized to the continuous case, will give us Hamiltonian flow.

\section{Numeric labels}

We now focus on labels that can be identified by numbers, where we have a bijective map between the label and a number in a set. The definitions in the previous section readily apply for labels identified by integers. Let $z \in \mathbbm{Z}$, we have $\mathbbm{c}=\sum\limits_{z \in \mathbbm{Z}} D(z) \mathbbm{s}(z)$ and the deterministic map becomes $f:\mathbbm{Z} \leftrightarrow \mathbbm{Z}$.

For the continuous case, one may simply expect to replace $z \in \mathbbm{Z}$ with $r \in \mathbbm{R}$, but this does not work. In the continuous limit, we would have $\mathbbm{c}=\int\limits_{r \in \mathbbm{R}} \rho(r) dr \mathbbm{s}(r)$, where $\rho(r) = D(r) / dr$. The continuous distribution $\rho$ is a density, defined over interval $dr$. That is: it's really $\rho(r, dr)$, function of both the center and the width of the interval. A bijective map on just $r$ is not sufficient, $dr$ must be mapped as well. On one side we claim the state fully identified by $r$, on the other $dr$ (a different label) is required for the density and the bijective map. We can't have it both ways.

In the continuous case, then, the appropriate label corresponds to a cell, not a point. This also makes physical sense, as we never deal with points per se, but widths that can be made arbitrarily small. A cell of $\mathbbm{R}$ corresponds to $\mathbbm{R}^2$: a center and a width.\footnote{In other words: the particles of the system are not point-like, but infinitesimal cell-like. Not only does this lead to a more direct understanding of classical phase space, it is also more consistent with general relativity (the mass is spread across a small region, not forming a singularity) and quantum mechanics (position is always the central value of a small distribution).} With this in mind, we have the following definitions.

\begin{defn}\label{sdof}
A \emph{degree of freedom} is a state variable identified by an infinitesimal cell of a one dimensional manifold $\mathbbm{Q}$.\footnote{For each label $i$ there is one and only one infinitesimal cell.}
\end{defn}

\begin{defn}\label{generalizedCoordinates}
We define the \emph{generalized coordinate} $q$ as the center of each cell. We define the \emph{cell number}\footnote{This represents the classical analogue of the wave number.} $k$ such that $k \, dq$ represents the width of each cell.
\end{defn}

\begin{cor}\label{continuousLabels}
Each degree of freedom is identified by the pair of labels $\langle q,k \rangle \in \mathbf{T}^*\mathbbm{Q}$ where $\mathbf{T}^*\mathbbm{Q}$ is the cotangent bundle of $\mathbbm{Q}$. Under coordinate changes, $k\,dq$ and $dq\wedge dk$ are invariant and $k$ is contravariant.
\end{cor}

Let's start with the discrete case. Let $Q$ be a finite region of $\mathbbm{Q}$. Divide the region into $N$ equal intervals of center $q$ and length $\Delta q$.\footnote{The definition could be mathematically more general. We use equally spaced intervals as it makes the derivation less cumbersome.} Let $K$ be a finite region of $\mathbbm{R}$. Divide the region into $M$ equal intervals of center $k$ and length $\Delta k$. Let $I$ be the label set identified by the cells in $Q$ with center $q(i)$ and width $k(i) \Delta q$. The configuration state will be given by $\mathbbm{c}=\sum\limits_{i \in I} \rho(q(i), k(i)) \Delta q \Delta k \, \mathbbm{s}(q(i), k(i))$, where $\rho(q(i), k(i)) = D(q(i), k(i)) / (\Delta q \Delta k)$.

As we increase $N$ and $M$, the cardinality of the label set $I$ increases, $\Delta q$ and $\Delta k$ decrease and so does the width of the cells. Our definitions, though, do not change: for each $\langle q,k \rangle$ we have one and only one cell, one state. In the limit, we will cover every possible center $q$ and every possible cell number $k$. Our configuration state becomes $\mathbbm{c}=\int\limits_{q \in Q \; k\in K} \rho(q, k) dq \wedge dk \, \mathbbm{s}(q, k)$. We can repeat the process increasing or changing the region covered by $Q$ and $K$, until we cover all of $\mathbbm{Q}$ and $\mathbbm{R}$.

If we apply a coordinate change $q'=q'(q)$, we change the labels but the state must remain the same, defined at the same point with the same width.\footnote{The notion that states are what they are no matter how we label them already contains the principle of relativity.} $k' dq'$ must be then equal to $k dq$; $k'= k dq / dq'$ is contravariant, $dq' \wedge dk'= (dq'/dq) \, (dq \wedge dk) \, (dq/dq') = dq \wedge dk$ is invariant. This means that $\rho$ is also invariant, which makes sense: the density depends only on the state (the cell), and not the coordinate chosen to represent it. As $k$ is contravariant, in the limit $K$ becomes the cotangent space of $Q$. Phase space is the cotangent bundle $\mathbf{T}^*\mathbbm{Q}$.

\begin{defn}\label{relativeCardinality}
Let $I \subset \mathbf{T}^*\mathbbm{Q}$ be a closed dense set of labels, a subset of a degree of freedom (i.e. a region of phase space). We define \emph{relative cardinality} $n(I) \rightarrow \mathbbm{R}$ as the ratio between the number of labels in $I$ and the ones of a reference set $I_0$. We find $n(I)=\int \omega$, where $\omega = \hbar \, dq \wedge dk$ and $\hbar$ is the constant that determines the unit (i.e. fixed so that $n(I_0) = 1$).
\end{defn}

As we deal with continuous labels, a label set is uncountable: given any range $\delta q$ and $\delta k$ there are an infinite number of possible labels $I$. Let $I_0$ be another set, defined on $\delta q_0$ and $\delta k_0$. In the discrete case, we have:

\begin{align*}
\frac{n(I)}{n(I_0)} = \frac{\delta q \delta k / \Delta q \Delta k}{\delta q_0 \delta k_0 / \Delta q \Delta k} = \frac{\delta q \delta k}{\delta q_0 \delta k_0}
\end{align*}

That is: while the cardinality of each set diverges, the ratio between the two remains finite.\footnote{Such treatment is equivalent to what one does in information theory to extend Shannon's entropy\cite{Shannon} to the continuous case. There is a link between label cardinality and informational entropy, including Jaynes' invariant formulation\cite{Jaynes}, which we do not explore here for brevity.} For continuous state variables we can define the relative cardinality as the ratio of a label set and our reference set $I_0$. That is $n(I)=\hbar \delta q \delta k$ where $\hbar=n(I_0)/(\delta q_0 \delta k_0)$ and $n(I_0)\equiv1$ by definition.\footnote{The choice of angular momentum for $\hbar$ derives from the relationship between $p$ and $dq/dt$, which will be derived later. The value for $\hbar$ in classical mechanics is arbitrary: we have no physical reason to choose a reference label set over another. In quantum mechanics, instead, we will have a preference: the set that corresponds to a single quantized system. We use this arbitrariness to set the value of our classical analogue of $\hbar$ to the known constant.} In general, the label set is not rectangular in $\langle q,k \rangle$ and we have $n(I)=\int \hbar dq \wedge dk$.\footnote{It should be stressed that no notion of uncertainty is used here. Though there is a link between label cardinality and uncertainty, which we do not explore here for brevity, label cardinality is the more fundamental quantity as we can still count the number of possible states even in an ideal case of no uncertainty.}

\begin{defn}\label{conjugateMomentum}
We define the \emph{conjugate momentum} as $p=\hbar k$.
\end{defn}

As $\hbar dq \wedge dk$ is truly fundamental, it is customary to group $\hbar$ with $k$: $dq \wedge (\hbar dk) = dq \wedge dp$. This way the area formed by generalized coordinate and conjugate momentum corresponds to the cardinality of states defined on such area.

\begin{cor}\label{continuousConjugateRelationships}
Each degree of freedom is identified by the pair of labels $\langle q,p \rangle \in \mathbf{T}^*\mathbbm{Q}$. $\theta_0 = p dq$ and $\omega = dq \wedge dp$ are invariant; $p$ is contravariant. The configuration state is $\mathbbm{c}=\int \rho(q,p) dq \wedge dp \, \mathbbm{s}(q, p)$, where $\rho(q,p)\equiv D(q,p) / \omega$ is the distribution density for each label.
\end{cor}

This restates all the previous findings in terms of conjugate momentum. In the language of differential geometry, we recognize phase space as the cotangent bundle $\mathbf{T}^*\mathbbm{Q}$, the function $\rho$, the one-form $\theta_0$ and the two-form $\omega$. They are truly fundamental objects as they are intimately linked to the way states are defined ($\rho$ is the distribution, $\theta_0 / \hbar$ is the cell width) and counted ($\omega$ is the cardinality of the labels) and they don't depend on coordinate choice.

As we'll need to work with different physical units, which may change $q$ or $p$ independently, it is useful to introduce the following notation and concepts.

\begin{defn}\label{differentialCardinality}
Let $dq$ be a finite contiguous range for a continuous state variable, we define $n_q(dq)$ as the cardinality of the set $I_{dq}$ that comprises all the labels in the range. For contravariant (conjugate) variables, we reverse the indices in the notation: $n^p(dp)$.
\end{defn}

\begin{cor}\label{linearCardinality}
Given $dq = a dq^a + b dq^b$, then $n_q(dq) = a n_q(dq^a) + b n_q(dq^b)$. That is, the relative cardinality is a linear operator.
\end{cor}

The linearity should be obvious: as we combine or increase the range, we combine or increase the number of possible labels.\footnote{Since the relative cardinality is linear, invariant under coordinate transformation and conserved by deterministic and reversible motion, it should come as no surprise that linearity will play a significant role. This is true for both classical (e.g. forces combine linearly) and quantum (e.g. linear inner product) concepts that are tightly linked to this fundamental operator.}

\begin{defn}\label{cardinalityConversionConstant}
Let $q^a$ and $q^b$ be two continuous state variables measured in different units, and $n_a(dq^a)$ and $n_b(dq^b)$ the relative cardinality defined on intervals of the respective units. We define the \emph{cardinality conversion constant} $\mathfrak{C}^b_a$ that converts a range of $q^a$ to a range of $q^b$ with the same cardinality. That is: $dq^b = \mathfrak{C}^b_a dq^a$ and $n_b(dq^b)=n_a(dq^a)$. For contravariant (conjugate) variables, we reverse the indices in the notation: $dp_b = \mathfrak{C}^a_b dp_a$.
\end{defn}

As the relative cardinality $n_q(dq)$ is invariant, the function $n_q$ cannot be since $dq$ is covariant. This means that, when changing units, we have to be careful.\footnote{This is only a problem when considering the number of labels of a single state variable $n_q(dq)$, instead of a whole degree of freedom $n_{\langle q, p \rangle}(\langle dq, dp \rangle)$: $\langle dq, dp \rangle$ is invariant and so is $n_{\langle q, p \rangle}$.}
Fortunately, since the $n_q$ is linear, we can always find a coefficient that allows us to convert. For example: $\hbar \equiv \mathfrak{C}^k_p$ is the cardinality conversion constant between cell number and conjugate momentum. In the future, we will have to convert intervals of time into space, or intervals of velocity into conjugate momentum, so the idea of cardinality conversion constants will reappear.

\section{Single degree of freedom}

Now that we have properly defined continuous labels and the relative cardinality of their sets, we will extend bijective maps to the continuous case. As these must preserve the cardinality of label sets, they are constrained in their form. The use of a bijective map for infinitesimal time evolution will lead us to Hamilton's equations. In this section we'll study a single degree of freedom, then extend in the next two sections to multiple degrees of freedom and to the time dependent case.

\begin{defn}\label{canonical}
We call \emph{canonical transformation} the continuous limit of a bijective map, as defined by \ref{detrevMap}, on $\mathbf{T}^*\mathbbm{Q}$.
\end{defn}

\begin{prop}\label{continuousMapping}
A canonical transformation must be continuous and preserve $\omega$.
\end{prop}

As we saw in \ref{labelsCount}, a bijective map conserves the cardinality of labels. In the continuous case, it will conserve relative cardinality and therefore $\omega$. The map must be continuous in $q$: suppose it isn't, it would split some cells into two parts, a cell would not be mapped to one and only one other cell, the mapping would not be bijective. The reverse mapping must be continuous in $q$ as well, or the inverse would not map to one and only one cell.
\begin{align*}
dq' &= \frac{\partial q'}{\partial q} dq + \frac{\partial q'}{\partial p} dp \\
dq &= \frac{\partial q}{\partial q'} dq' + \frac{\partial q}{\partial p'} dp'
\end{align*}
We can re-express $dp'$ in terms of $dq$ and $dp$, as the conservation of $\omega$ means the map is non-degenerate. All partial derivatives are well defined, and therefore the mapping is continuous in $p$ as well.

\begin{cor}\label{sdofInvariant}
Let $v$ and $w$ be two vectors defined on the tangent space of the phase space $\mathbf{T}^*\mathbbm{Q}$ for one degree of freedom. Let
\begin{align*}
\omega_{ab} = \left[
  \begin{array}{cc}
    0 & 1 \\
    -1 & 0 \\
  \end{array}
\right] \\
\end{align*}
then $v'^{a} \omega_{ab} w'^{b}=v^{a} \omega_{ab} w^{b}$ under a canonical transformation, where $a$ and $b$ represent components along coordinates $\xi^{a} \equiv \{q,p\}$.
\end{cor}

Here we are simply expressing $\omega$, $v$ and $w$ in terms of their respective components and underlining the fact that $\omega$ defines the metric conserved under canonical transformations.

\begin{lem}\label{genAntisim}
Let $v$ and $w$ be two vectors. Let $v^{a} \omega_{ab} w^{b}$ be an antisymmetric product conserved under a continuous transformation parameterized by $t$. We can then define a function $H$ such that given $S^{a} \equiv d_{t}\xi^{a}$ and $S_{b} \equiv S^{a} \omega_{ab}$, we have $S_{a} = \partial_{a}H$.
\end{lem}

$S^{a}$ is the vector field that represents how the state variables change. Simply applying the vector transformation rules under continuous transformation we have:
\begin{align*}
v^{a} \omega_{ab} w^{b} &= v'^{a} \omega_{ab} w'^{b}  \\
&= (v^{a} + \partial_{c} S^{a} v^{c} dt) \omega_{ab} ( w^{b} + \partial_{d} S^{b} w^{d} dt) \\
&= v^{a} \omega_{ab} w^{b} + (\partial_{c} S^{a} v^{c} \omega_{ab} w^{b} \\
 &+ v^{a} \omega_{ab} \partial_{d} S^{b} w^{d}) dt + O(dt^2)
\end{align*}
\begin{align*}
v^{c} w^{b} \partial_{c} S_{b} - v^{a} w^{d} \partial_{d} S_{a} = 0
\end{align*}
\begin{align*}
\partial_{a} S_{b} - \partial_{b} S_{a} &= curl(S_{a}) = 0 \\
S_{a} &= \partial_{a}H
\end{align*}

\begin{prop}\label{sdofHam}
The time evolution for a single degree of freedom is given by:
\begin{align*}
d_{t}q &= \partial_{p} H \\
d_{t}p &= - \partial_{q} H
\end{align*}
\end{prop}

Simply expand \ref{genAntisim} with the metric defined in \ref{sdofInvariant}. We recognize Hamilton's equations for one degree of freedom\cite{classical_dynamics}.

\section{Multiple independent degrees of freedom}

\begin{prop}\label{mdofInvariant}
Let $v$ and $w$ be two vectors defined on the tangent space of the phase space $\mathbf{T}^*\mathbbm{Q}$ for two independent degrees of freedom. Let $a$ and $b$ be indices for the state variables $\xi^a\equiv \{q^i, p_i\}$. Let
\begin{align*}
\omega_{ab} =  \left[
  \begin{array}{cc}
    0 & 1 \\
    -1 & 0 \\
  \end{array}
\right] \otimes \left[
  \begin{array}{cc}
    1 & 0 \\
    0 & 1 \\
  \end{array}
\right] =
\left[
  \begin{array}{cccc}
    0 & 0 & 1 & 0 \\
    0 & 0 & 0 & 1 \\
    -1 & 0 & 0 & 0 \\
    0 & -1 & 0 & 0 \\
  \end{array}
\right] \\
\end{align*}
then $v'^{a} \omega_{a b} w'^{b}=v^{a} \omega_{ab} w^{b}$ under a canonical transformation.
\end{prop}

The independence between degrees of freedom corresponds to orthogonality in phase space: from \ref{labelsCombine} the product between the number of labels on each d.o.f. (i.e. the area), must be equal to the number of combined labels (i.e. the hyper-volume), which is true only if the d.o.f are orthogonal in phase space. From \ref{labelsCount}, the mapping will preserve the cardinality of labels, the area\footnote{We assume we are using the same unit across d.o.f.} on each d.o.f, and the independence, orthogonality across d.o.f.\footnote{These statements provide a direct physical interpretation for Gromov's non-squeezing theorem\cite{Gromov,deGosson,Stewart}.} This is equivalent to requiring the conservation of the scalar product across independent degrees of freedom, while still requiring conservation of the vector product within. That leads us to the metric defined by \ref{mdofInvariant}.
The metric generalizes \ref{sdofInvariant} to give us the cardinality of labels defined on the area given by two arbitrary directions in phase space. For an infinitesimal region, this corresponds to $dq^1 \wedge dp_1 + dq^2 \wedge dp_2$, the sum of the projections on the independent planes. Moreover, volume in phase space corresponds to the cardinality of the combined labels (i.e. the states), and is therefore conserved: this is Liouville's theorem for Hamiltonian mechanics.

\begin{prop}\label{mdofHam}
The evolution for multiple degrees of freedom is given by:
\begin{align*}
d_{t}q^i &= \partial_{p_i} H \\
d_{t}p_i &= - \partial_{q^i} H
\end{align*}
\end{prop}

Expand \ref{genAntisim} with the metric defined in \ref{mdofInvariant}. We recognize Hamilton's equations for multiple degrees of freedom\cite{classical_dynamics}.

\section{Time dependence}

So far we have assumed that neither state labeling nor mapping changes in time. If they do, we also need to to use time as a label and therefore introduce an extra degree of freedom.

\begin{defn}\label{tdof}
We call \emph{extended phase space} the cotangent bundle $\mathbf{T}^*\mathcal{M}$, where the manifold $\mathcal{M}$ identifies the possible center values for all infinitesimal cells at all times. We call \emph{temporal degree of freedom} the state variable identified by temporal cells. The center of each cell is identified by $t$, the width by $\hat{\omega} dt$, the conjugate variable $E\equiv\hbar\hat{\omega}$.\footnote{$\hat{\omega}$ is the classical analogue of the wave frequency. We use $\hat{\omega}$ to distinguish from the phase-space metric $\omega$.}
\end{defn}

\begin{prop}\label{tdofMonotonic}
Let $s$ be the parameter of a trajectory in the extended phase space of a deterministic and reversible system. The trajectory must be continuous. There must exist a strictly monotonic function $t(s)$.
\end{prop}

The trajectory has to be continuous in both standard and temporal variables because of \ref{continuousMapping}. Since determinism and reversibility are defined in time, the trajectory must traverse all times once and only once: we must have an invertible mapping between $t$ and $s$, which means we must have a strictly monotonic $t(s)$.

\begin{defn}\label{tdofAntistates}
We call \emph{standard states} those connected by a trajectory where $d_{s}t>0$. We call \emph{anti-states} those connected by a trajectory where $d_{s}t<0$.
\end{defn}

Since $t(s)$ is strictly monotonic, $d_{s}t$ along a trajectory cannot change sign, so we have the division between standard and anti-states. Note that since the parametrization is conventional and can be changed to $s'=-s$, what we call standard and anti-states is also conventional. What is physical and not conventional, though, is that standard and anti-states cannot be connected by deterministic and reversible evolution.\footnote{This represents a classical analogue for quantum anti-particle states.}

\begin{prop}\label{tdofInvariant}
Let $v$ and $w$ be two vectors defined on the tangent space of extended phase space $\mathbf{T}^*\mathcal{M}$ for the temporal degree of freedom and one standard degree of freedom. Let $a$ and $b$ be indices for the state variables $\xi^a\equiv\{t, q, E, p\}$. Let
\begin{align*}
\omega_{ab} =  \left[
  \begin{array}{cc}
    0 & 1 \\
    -1 & 0 \\
  \end{array}
\right] \otimes \left[
  \begin{array}{cc}
    -1 & 0 \\
    0 & 1 \\
  \end{array}
\right]
= \left[
  \begin{array}{cccc}
    0 & 0 & -1 & 0 \\
    0 & 0 & 0 & 1 \\
    1 & 0 & 0 & 0 \\
    0 & -1 & 0 & 0 \\
  \end{array}
\right] \\
\end{align*}
then $v'^{a} \omega_{ab} w'^{b}=v^{a} \omega_{ab} w^{b}$ under deterministic and reversible evolution.
\end{prop}

$\langle t, E \rangle$ are not independent from $\langle q, p \rangle$ as they do not define new states. So they are not necessarily orthogonal in the extended phase space. States are defined on the plane where $\langle q, p \rangle$ (maximally) change: this is not the plane of constant $\langle t, E \rangle$ (they are not orthogonal) where $dq \wedge dp$ is defined, but the plane perpendicular to constant $\langle q, p \rangle$ where $dt \wedge dE$ is defined. On that plane we can properly count states and define our metric.

We have a right triangle-like relationship between the plane where the metric is defined and its projections on the planes defined by each d.o.f., similar to the multiple d.o.f.:
\begin{align*}
m.d.o.f \;\;\; &dq^1 \wedge dp_1 + dq^2 \wedge dp_2 = \omega \\
t.d.o.f \;\;\; &dt \wedge dE + \omega = dq \wedge dp \\
\end{align*}
But in the previous case, the right angle was between the two independent d.o.f.. In this case, the right angle is between the metric and the plane of constant $\langle q, p \rangle$ where $dt \wedge dE$ is defined. We rewrite it as $dq \wedge dp - dt \wedge dE = \omega$. This corresponds to the Minkowski product across d.o.f. and the vector product within. The metric \ref{tdofInvariant}, with a space-like convention, still gives us the cardinality of labels within a degree of freedom, adjusting \ref{sdofInvariant} to avoid ``double counting''.

\begin{prop}\label{tdofHam}
The evolution for time varying multiple degrees of freedom is given by:
\begin{align*}
d_{s}t &= - \partial_{E} \mathcal{H} \\
d_{s}E &= \partial_{t} \mathcal{H} \\
d_{s}q^i &= \partial_{p_i} \mathcal{H} \\
d_{s}p_i &= - \partial_{q^i} \mathcal{H}
\end{align*}
\end{prop}

Take the metric from \ref{tdofInvariant}, add multiple independent d.o.f as in \ref{mdofInvariant}, use \ref{genAntisim} with the parameter $s$ instead of $t$ and generator $\mathcal{H}$ instead of $H$.

If we set\footnote{We avoided using $p_{n+1}$ as it hides the minus sign from the metric, making it seem that the temporal d.o.f is just another independent d.o.f.} $q^{n+1}=t$ and $p_{n+1}=-E$, we recognize Hamilton's equations in the extended phase space\footnote{As in Struckmeier\cite{Struckmeier}, $d_{s}t$ need not be unitary.}\cite{Synge,Lanczos}.

\begin{prop}\label{tdofConstrain}
The evolution is constrained by $\mathcal{H}=k$.
\end{prop}

Since $\mathcal{H}$ is constant through the evolution, it can serve both as the generating function and as the evolution constraint. By convention, we can set $\mathcal{H}=0$ without loss of generality as changing $\mathcal{H}$ by a constant does not change the equation of motion. This reduces extended phase space to $2N + 1$ components, the state variables plus time.

\section{Kinematics}
In the previous section we made no constraint on what our state variables $q^i$ actually represent. In this section we turn our attention to the study of the motion of a body. In particular, we will introduce another assumption: that the trajectories are enough to fully describe the system. We can expect this to hold true if the system is elementary (it has no relevant internal structure) and is sufficiently isolated. At that point, if the motion is the result of a deterministic and reversible process, no two trajectories can be attributed to the same state as there is nothing else that could affect them. As position and time will be state variables, the space-time manifold will coincide with $\mathcal{M}$ and the extended phase space with $\mathbf{T}^*\mathcal{M}$.

\begin{assump3}\label{kinematicAssumption}
The study of the trajectory (kinematics) of a body is equivalent to study of its state (dynamics) under deterministic and reversible evolution.
\end{assump3}

\begin{cor}\label{singleTrajectory}
Fix a system under deterministic and reversible evolution. Given all its possible trajectories $x^\alpha(s)$ in the space-time manifold $\mathcal{M}$ and all its possible trajectories $\xi^a(s)$ in the extended phase space $\mathbf{T}^*\mathcal{M}$, there exists a bijective function $f: x^\alpha(s) \leftrightarrow \xi^a(s)$ that links each space-time trajectory with one and only one phase-space trajectory.
\end{cor}

If studying the motion and state evolution are equivalent, then we must be able to go back and forth between the two pictures. Without losing generality, we can choose $x^\alpha=f(t,q^i)$ to be a linear function of only $q^i$ and $t$.

\begin{defn}\label{invariantSpeed}
Let $x$ be a space variable and $t$ the time variable. We define the \emph{invariant speed} as the cardinality conversion constant $c \equiv \mathfrak{C}^x_t$ between $x$ and $t$. That is, if $dx=cdt$ then $n_x(dx)=n_t(dt)$.
\end{defn}

\begin{prop}\label{locallyMinkowski}
The space-time manifold $\mathcal{M}$ is a locally Minkowskian Riemannian manifold. That is, there exists a metric $g$ that at any point P can be expressed, with a suitable choice of coordinate, as $g=dx^\alpha g_{\alpha \beta}(P) dx^\beta=dx^\alpha\eta_{\alpha \beta}dx^\beta=(dx^i)^2 - (dx^0)^2=(dx^i)^2 - c^2dt^2$, where $\eta_{\alpha \beta}$ is the Minkowski metric.
\end{prop}

As states are defined on intervals, a metric $g$ must be defined on $\mathcal{M}$. Such a metric must be consistent with $\omega$ as defined on $\mathbf{T}^*\mathcal{M}$, as both must be invariant under coordinate transformations. The idea is that each $(dx^\alpha)^2$ can be made to represent both a length squared in space-time and an area in phase space, linking the two metrics. For each spatial d.o.f. fix $dp_i=\lambda dq^i$, we have $dq^i \wedge dp_i = \lambda (dq^i)^2$ where $\lambda$ converts from length squared to the label cardinality contained in the area. For the temporal d.o.f fix $dE = \lambda c^2 dt$, we have $dt \wedge dE = \lambda c^2 dt^2$. That is: the area in time squared is converted to an area in length squared that has the same density of states (as per definition of $c$) and then is converted again to its label cardinality. The phase-space metric is $\omega = \lambda [(dq^i)^2 - c^2 dt^2]$.

Assume the choice of coordinates $x^\alpha$ locally diagonalizes $g(x^\alpha)$, each diagonal element being either $\pm 1$ (such coordinate system always exists). Set $x^i=q^i$ and $x^0=ct$. We have $\omega = \lambda ((dx^i)^2 - (dx^0)^2)$ and $g=dx^\alpha g_{\alpha \beta} dx^\beta=dx^\alpha g_{\alpha \alpha}dx^\alpha$. Both are invariant under coordinate transformation for any $dx^\alpha$. This can only be if $g_{ii}(x^\alpha)=1$ and $g_{00}(x^\alpha)=-1$.

It is fitting that deterministic and reversible evolution requires space-time to be locally Minkowskian, as this clearly defines past and future events. To make us understand better the role of $c$, as we defined it, we prove the following.

\begin{prop}\label{speedLimit}
The speed of a body cannot exceed the invariant speed $c$ under the kinematic equivalence assumption.
\end{prop}

Consider a movement $ds$ along any trajectory. This will go through $|n_t(dt)|$ labels in time and $|n_x(dx)|$ labels in space. As we go through labels in time, we may go through fewer labels in space (e.g. the particle remains still) but not more: we cannot skip space labels or the motion would not be continuous. So we have:
\begin{align*}
\frac{|n_x(dx)|}{|n_t(dt)|} &= \frac{|n_x(dx)|}{|n_x(c dt)|} = \frac{|dx|}{|c dt|} \leq 1 \\
\frac{|dx|}{|dt|} & \leq c \\
\end{align*}

Thus we find that $c$ is the well known relativistic constant.\footnote{Deterministic and reversible motion at speeds greater than $c$ (i.e. a tachyon) is not allowed. Non-deterministic and non-reversible motion at speeds greater than $c$ (e.g. correlations at a distance) is allowed.}

\begin{prop}\label{initialConditions}
Let $x^\alpha=\{ct, q^i\}$ and $u^\alpha = d_s x^\alpha$ be the four-velocity, where the parametrization $s$ is chosen, by convention, to be proper time. The position $x^\alpha$ and velocity $u^\alpha$ are necessary and sufficient initial conditions to determine the state of the system and its whole trajectory.
\end{prop}

As the equations of motion \ref{tdofHam} can be at most second order in $\{t, q^i\}$, they can at most be second order in $x^\alpha=\{ct, q^i\}$ so only position and velocity can be candidates for the initial conditions. Fixing time, phase space has $2N$ components, too big to be covered by position only, but just right to be covered by both.

\begin{defn}\label{inertialMass}
Let $p_\alpha$ be a component of conjugate momentum and $u_\alpha \equiv g_{\alpha \beta}u^\beta$ be a contravariant component of the four velocity. We define the \emph{inertial mass} as the cardinality conversion constant $m \equiv \mathfrak{C}^p_u$ between $p$ and $u$. That is, if $dp_\alpha=mdu_\alpha$ then $n^p(dp_\alpha)=n^u(du_\alpha)$.\footnote{Note that the kinematic assumption rules out massless particles: they can have the same position and velocity with different momentum, therefore the same trajectory for different states. When generalizing to field theory, the derivative of the position will be replaced by the derivative of the field (i.e. its oscillation in space-time), which holds for massive and massless fields.}
\end{defn}

\begin{prop}\label{kineticMomentum}
Let $x^\alpha=\{ct, q^i\}$ and $p_\alpha=\{-E/c, p_i\}$. Then $p_\alpha= m g_{\alpha \beta}u^\beta + \hat{p}_\alpha(x)$, where $\hat{p}_\alpha : \mathcal{M} \rightarrow \mathbbm{R}$ is a function defined on the space-time manifold.
\end{prop}

Given \ref{initialConditions}, there must exist $p_\alpha=p_\alpha(x^\beta , u^\gamma)$. We express it in terms of $u_\alpha\equiv g_{\alpha \beta} u^\beta$. We have:
\begin{align*}
\omega &= q^i\wedge p_i - ct \wedge E/c \\
&=x^\alpha \wedge p_\alpha \\
&=dx^\alpha \wedge \frac{\partial p_\alpha}{\partial u_\beta}du_\beta + dx^\alpha \wedge \frac{\partial p_\alpha}{\partial x^\gamma}dx^\gamma \\
&=\frac{\partial p_\alpha}{\partial u_\beta}du_\beta dx^\alpha \\
\end{align*}
Consider the expression $m du_\alpha dx^\alpha$: this invariant gives us the density of labels in position and velocity converted to the conjugate variables. That is: $\omega=m du_\alpha dx^\alpha$. Combining the two:
\begin{align*}
\frac{\partial p_\alpha}{\partial u_\beta} &= m \delta^\beta_\alpha \\
p_\alpha &= m g_{\alpha \beta}u^\beta + \hat{p}_\alpha(x)
\end{align*}
where $\hat{p}$ is an arbitrary function.\footnote{The conjugate momentum $p$ and the kinetic momentum $mu$ are therefore different in general: they are linked only through their differentials. Moreover, $p$ is unique only once $\hat{p}$ is fixed, but $\hat{p}$ is arbitrary. Given that $u$ determines the trajectory, any transformation in $\{p, \hat{p} \}$ that does not change $u$ describes the same motion. We call \emph{gauge transformation} such a transformation that redefines conjugate momentum while keeping kinetic momentum unvaried. This is the particle mechanics equivalent of gauge transformations in field theory.}

To convince ourselves that $m$ is indeed the inertial mass, consider applying a force. We are changing the state through the velocity, meaning changing the conjugate momentum. The higher the mass, the more states we'll have to go through to reach the same velocity. The higher the mass, the more change is required, the more force needs to be applied.

\begin{prop}\label{kineticHamiltonian}
The extended Hamiltonian is $\mathcal{H}=\frac{1}{2m}(p_\alpha-\hat{p}_\alpha(x))g^{\alpha\beta}(p_\beta-\hat{p}_\beta(x))+\hat{\mathcal{H}}(x)$, where $\hat{\mathcal{H}} : \mathcal{M} \rightarrow \mathbbm{R}$ is a function defined on the space-time manifold.
\end{prop}
We have:
\begin{align*}
\frac{dx^\alpha}{ds} &= u^\alpha \\
&= \frac{1}{m}g^{\alpha\beta}(p_\beta-\hat{p}_\beta) \\
&= \frac{\partial \mathcal{H}}{\partial p_\alpha} \\
\end{align*}
Integrating we have the expression for the Hamiltonian, where $\hat{\mathcal{H}}$ is an arbitrary function.

Now that we have found the constrained form of the Hamiltonian, we show that this is compatible with the established fundamental classical theories.

\begin{prop}\label{relativisticEM}
Let $\hat{p}_\alpha = q A_\alpha$, $F_{\alpha \beta} \equiv \partial_\alpha A_\beta - \partial_\beta A_\alpha$ and $\hat{\mathcal{H}} = 0$. Then the equations of motion are $m \nabla_u u^\alpha = g^{\alpha\beta} F_{\beta \gamma} q u^\gamma$. These are the relativistic equations for a charged particle.
\end{prop}

We first derive the following relationship for later use:
\begin{align*}
\partial_\alpha \delta^\beta_\gamma &= 0 = \partial_\alpha g^{\beta\delta} g_{\delta\gamma} + g^{\beta\delta} \partial_\alpha g_{\delta\gamma}\\
\partial_\alpha g^{\beta\epsilon} &= - g^{\beta\delta} g^{\gamma\epsilon} \partial_\alpha g_{\delta\gamma}
\end{align*}

We expand \ref{tdofHam} using \ref{kineticHamiltonian}:
\begin{align*}
u^\alpha &= \frac{dx^\alpha}{ds} = \frac{\partial \mathcal{H}}{dp_\alpha} \\
&= \frac{1}{m}g^{\alpha\beta}(p_\beta-\hat{p}_\beta) \\
d_s p_\alpha &= - \frac{\partial \mathcal{H}}{\partial q^\alpha} \\
&=\frac{1}{2m}[\partial_\alpha \hat{p}_\beta g^{\beta \gamma} (p_\gamma -\hat{p}_\gamma) \\
 &- (p_\beta -\hat{p}_\beta) \partial_\alpha g^{\beta \gamma} (p_\gamma -\hat{p}_\gamma) \\
 &+ (p_\beta -\hat{p}_\beta) g^{\beta \gamma} \partial_\alpha \hat{p}_\gamma ]- \partial_\alpha \hat{\mathcal{H}} \\
&=\frac{1}{2}[\partial_\alpha \hat{p}_\beta u^\beta
- m u^\delta g_{\delta\beta} \partial_\alpha g^{\beta \gamma} u^\epsilon g_{\epsilon\gamma}
+ u^\gamma \partial_\alpha \hat{p}_\gamma ]- \partial_\alpha \hat{\mathcal{H}}
\end{align*}

We then calculate the four-force:
\begin{align*}
m d_s u^\alpha &= d_s g^{\alpha\beta}(p_\beta-\hat{p}_\beta) + g^{\alpha\beta} d_s (p_\beta-\hat{p}_\beta) \\
&= \partial_\gamma g^{\alpha\beta} d_s x^\gamma m g_{\beta \delta} u^\delta + g^{\alpha\beta} (d_s p_\beta - \partial_\gamma \hat{p}_\beta d_s x^\gamma) \\
&= \partial_\gamma g^{\alpha\beta} u^\gamma m g_{\beta \delta} u^\delta + g^{\alpha\beta} \frac{1}{2} [\partial_\beta \hat{p}_\gamma u^\gamma \\
&- m u^\epsilon g_{\epsilon\gamma} \partial_\beta g^{\gamma \delta} u^\zeta g_{\zeta\delta} \\
&+ u^\delta \partial_\beta \hat{p}_\delta ]- g^{\alpha\beta} \partial_\beta \hat{\mathcal{H}} - g^{\alpha\beta} \partial_\gamma \hat{p}_\beta u^\gamma \\
&= m  g_{\beta \delta} \partial_\gamma g^{\alpha\beta} u^\gamma u^\delta - \frac{1}{2} m g^{\alpha\beta} g_{\zeta\delta} g_{\epsilon\gamma} \partial_\beta g^{\gamma \delta} u^\epsilon u^\zeta  \\
&+ g^{\alpha\beta} \partial_\beta \hat{p}_\gamma u^\gamma - g^{\alpha\beta} \partial_\gamma \hat{p}_\beta u^\gamma
- g^{\alpha\beta} \partial_\beta \hat{\mathcal{H}}\\
&= - m  g^{\alpha \beta} \partial_\gamma g_{\beta\delta} u^\gamma u^\delta + \frac{1}{2} m g^{\alpha\beta} \partial_\beta g_{\gamma \delta} u^\gamma u^\delta  \\
&+ g^{\alpha\beta} (\partial_\beta \hat{p}_\gamma - \partial_\gamma \hat{p}_\beta ) u^\gamma
- g^{\alpha\beta} \partial_\beta \hat{\mathcal{H}}\\
&= - m \frac{1}{2} g^{\alpha \beta} ( \partial_\gamma g_{\beta\delta} + \partial_\delta g_{\beta\gamma} - \partial_\beta g_{\gamma \delta} ) u^\gamma u^\delta  \\
&+ g^{\alpha\beta} (\partial_\beta \hat{p}_\gamma - \partial_\gamma \hat{p}_\beta ) u^\gamma
- g^{\alpha\beta} \partial_\beta \hat{\mathcal{H}}\\
&= - m \Gamma ^\alpha_{\ \gamma \delta} u^\gamma u^\delta + g^{\alpha\beta} (\partial_\beta \hat{p}_\gamma - \partial_\gamma \hat{p}_\beta ) u^\gamma
- g^{\alpha\beta} \partial_\beta \hat{\mathcal{H}}
\end{align*}
\begin{align*}
m \nabla_{u} u^\alpha &= m (d_s u^\alpha + \Gamma ^\alpha_{\ \gamma \delta} u^\gamma u^\delta)  \\
&= g^{\alpha\beta} (\partial_\beta \hat{p}_\gamma - \partial_\gamma \hat{p}_\beta ) u^\gamma - g^{\alpha\beta} \partial_\beta \hat{\mathcal{H}}\\
&= g^{\alpha\beta} (\nabla_\beta \hat{p}_\gamma - \nabla_\gamma \hat{p}_\beta ) u^\gamma - g^{\alpha\beta} \nabla_\beta \hat{\mathcal{H}}
\end{align*}
where $\nabla$ is the covariant derivative and $\Gamma$ the Christoffel symbols.

The equation is manifestly covariant. If $\hat{p}$ and $\hat{\mathcal{H}}$ are zero (i.e. in the absence of forces) we recognize the geodesic equation. We substitute $\hat{p}_\alpha = q A_\alpha$ and $\hat{\mathcal{H}} = 0$ and have:

\begin{align*}
m \nabla_{u} u^\alpha &= g^{\alpha\beta} (\partial_\beta A_\gamma - \partial_\gamma A_\beta ) q u^\gamma \\
&= g^{\alpha\beta} F_{\beta \gamma} q u^\gamma\\
\end{align*}

This is indeed compatible with general relativity and classical electromagnetism.

\begin{prop}\label{newtonianGravitation}
Assume $\mathcal{M}$ flat and time-independent motion. Then the Hamiltonian simplifies to $H=\frac{1}{2m}(p_i-\hat{p}_\alpha(x))^2+\hat{H}(x)$. Let $\hat{p}_i = 0$ and $\hat{H} = mV$, the equations of motion are $m d_t v^i = - m \partial_i V$, where $v^i=dx^i/dt$. These are equations for a particle under a Newtonian gravitational potential.
\end{prop}

Given flat space-time and time-independent motion, we can use $t$ as a parameter for the motion in $\mathbbm{Q}$. Repeating \ref{kineticMomentum} and \ref{initialConditions} for the non-relativistic case leads to the above Hamiltonian. Similar to \ref{relativisticEM}, use \ref{mdofHam} with the newly found Hamiltonian and find the equations of motion.

\begin{defn}\label{lagrangian}
Let $\mathcal{H}$ be an extended Hamiltonian defined on $\mathbf{T}^*\mathcal{M}$. Under the kinematic assumption, we can define the Legendre transform $\mathcal{L}=u^\alpha p_\alpha - \mathcal{H}$ which we call the \emph{extended Lagrangian}. For the time invariant case, we define the \emph{Lagrangian} $L=v^i p_i - H$.
\end{defn}

The Legendre transform can be defined only if $\mathcal{H}$ is convex in $p_\alpha$. The Hamiltonian found at \ref{kineticHamiltonian} is convex, so the Lagrangian can always be defined. While the kinematic equivalence assumption is sufficient, it's not necessary for the existence of a Lagrangian. What is necessary is \ref{initialConditions}: as long as position and velocity are enough to determine the state, $u^\alpha=f(q^\alpha, p_\alpha)$ must be monotonic in $p_\alpha$, which means $\mathcal{H}$ is convex in $p_\alpha$. What happens in that case is that the space-time trajectories may become more or less dense, $du_\alpha$ and $dx^\alpha$ would not be enough to define the label cardinality and therefore the state density $\rho$. In short: deterministic and reversible evolution gives us Hamiltonian mechanics, position and velocity as initial conditions gives us Lagrangian mechanics, and the kinematic assumption (which links the differentials of the state variables and the initial conditions) gives us the narrowed Hamiltonian form.

\section{Conclusion}

While there are many more details that we could have expanded upon, as we have touched many areas we could only scratch the surface. The intent is to give the overall picture, which could be summarized in the following points.

\begin{itemize}
  \item Classical states are those that describe every infinitesimal part.
  \item Classical particles are infinitesimal cells. Phase space is the cotangent bundle $\mathbf{T}^*\mathbbm{Q}$ because that's the space of infinitesimal cells.
  \item Hamiltonian mechanics coincides with deterministic and reversible evolution.
  \item Hamiltonian flow is the conservation of number of labels for each independent degree of freedom.
  \item Deterministic and reversible evolution is what ultimately gives space-time its locally Minkowskian nature.
  \item $\hbar$, $c$ and $m$ can be seen as conversion constants that preserve cardinality.
  \item Lagrangian mechanics coincides with position and velocity being necessary and sufficient initial conditions.
  \item The motion of an isolated elementary system, for which the kinematic assumption is valid, is restricted to a Hamiltonian that can describe fundamental classical forces.
\end{itemize}

This helps clarify and understand much of the classical framework in a more cohesive way and the notion of cardinality of labels and states is at the heart.

While this is limited to classical particle mechanics, it should be obvious how, at least in principle, this work could be extended. For field theories, the \emph{kinematic assumption} should be substituted by a \emph{kinematic field assumption}: what we are studying are not trajectories $x^\alpha(s)$ but fields $\psi(x)$. The field values at each point become a set of independent state variables, and their conjugates $\pi(x^\alpha)$ will be linked to $d\partial_s\psi(x^\alpha)$. For quantum mechanics, the \emph{infinite reducibility assumption} has to give way to an \emph{irreducibility assumption}: the state of parts of a quantum system cannot be known. The configuration state as a whole undergoes deterministic and reversible evolution, while the motion of its parts does not.

The hope is that, by continuing in this approach, we can shed more light on why the laws of physics are what they are; and show that they are not arbitrary rules, but necessary given few simple assumptions.

\section{Acknowledgements}

It is my pleasure to thank Christine A. Aidala, Lydia Bieri, Robert Geroch and Steven J. Miller for helpful discussions and support.


\begin{thebibliography}{0}

\bibitem{Shannon} Shannon, C. E., ``A mathematical theory of communication'', The Bell System Technical Journal, Vol. 27, pp. 379--423, 623--656, (1948).
\bibitem{Jaynes} Jaynes, E. T., ``Information theory and statistical mechanics'', Statistical Physics 3, pp. 181--218, (1963).
\bibitem{classical_dynamics} J. V. Jos\'{e}, E. J. Saletan, ``Classical Dynamics'', Cambridge University Press, (1998).
\bibitem{Gromov} Gromov, M. L., ``Pseudo holomorphic curves in symplectic manifolds'', Inventiones Mathematicae 82, pp. 307--347, (1985).
\bibitem{deGosson} de Gosson, M. A., ``The symplectic camel and the uncertainty principle: the tip of an iceberg?'', Foundations of Physics 39, pp. 194--214, (2009).
\bibitem{Stewart} Stewart, I., ``The symplectic camel'', Nature 329, pp. 17--18, (1987).
\bibitem{Lanczos} Lanczos, C., ``The variational principles of mechanics'', University of Toronto Press, (1949).
\bibitem{Synge} Synge, J. L., ``Classical dynamics'', Encyclopedia of Physics Vol 3/1, Springer (1960).
\bibitem{Struckmeier} Struckmeier, J., ``Hamiltonian dynamics on the symplectic extended phase space for autonomous and non-autonomous systems'', J. Phys. A: Math. Gen. 38, pp. 1257--1278, (2005).

\end{thebibliography}
\end{document}